\documentclass[a4paper,final]{article}
\usepackage[utf8]{inputenc}
\usepackage[english]{babel}
\usepackage{amsmath}

\usepackage{amsfonts}
\usepackage{amssymb}
\usepackage{graphicx}
\usepackage{setspace}
\usepackage{tikz}
\usepackage{hyperref}

\usepackage{algorithm}
\usepackage{algorithmic}
\renewcommand{\algorithmiccomment}[1]{\bgroup\hfill//~#1\egroup}

\usepackage{authblk}
\usepackage{xcolor}
\usepackage[normalem]{ulem}

\newtheorem{theorem}{Theorem}

\newtheorem{corollary}[theorem]{Corollary}

\newtheorem{definition}[theorem]{Definition}

\newtheorem{example}{Example}

\newtheorem{lemma}[theorem]{Lemma}

\newenvironment{proof}[1][Proof]{\textbf{#1.} }{\ \rule{0.5em}{0.5em}}

\title{Pareto optimal exchange with indifferent endowments}

\author[1]{Pavlos Eirinakis}
\author[2]{Ioannis Mourtos}
\author[2]{Michalis Samaris}
\affil[1]{Department of Industrial Management and Technology, University of Piraeus,\protect\\ \small Karaoli and Dimitriou 80, 18534, Piraeus, Greece, E-mail: {\tt pavlose@unipi.gr}}

\affil[2]{Department of Management Science and Technology, Athens University of Economics and Business,\protect\\ \small Athens, Greece, E-mail: {\tt mourtos@aueb.gr, michsam@aueb.gr}}

\begin{document}
\maketitle


\maketitle

\begin{abstract}
We investigate a market without money in which agents can offer certain goods (or multiple copies of an agent-specific good) in exchange for goods of other agents. The exchange must be balanced in the sense that each agent should receive a quantity of good(s) equal to the one she transfers to others. In addition, each agent has strict preferences over the agents from which she will receive goods, and there is an upper bound on the volume of each transaction and a weight reflecting its social importance or its cardinal utility for the two agents. We propose a simple variant of the Top Trading Cycles mechanism that finds a Pareto optimal balanced exchange. We then offer necessary and sufficient conditions for a balanced exchange to be Pareto optimal and exploit these to obtain a recognition procedure. This procedure can detect whether a given exchange is Pareto optimal and, if not, improve it to become Pareto optimal in polynomial time. Last, we show how to obtain a Pareto optimal balanced exchange of maximum weight in two special cases.

\end{abstract}

\section{Background and Contribution}


We are motivated by the growing interest in the sharing economy (e.g., \cite{Xu20}), along with allocation mechanisms without money \cite{Ashlagi16, Biro20} or schemes based on the so-called Barter economy \cite{Williams96}. We therefore assume a market where each agent is willing to provide a quantity of goods equal to the one she receives, hence the term \emph{balanced exchange}, introduced in \cite{Klijn17}. Regarding indivisible goods, this setting generalises kidney exchange \cite{Biro21} in the currently non-applicable scenario where the exchange is made among hospitals. Regarding divisible ones, our setting formalises the time-banking movement (see \cite{Andersson21} and references therein), in which people exchange services under the ``1 hour$=$1 hour" principle. Divisibility in our approach could also be considered appropriate for ``debt exchange", in which organisations or individuals could perform  debt reliefs of equal size alongside cycles of debtor-debtee pairs.

That is, the market under our consideration offers the exchange of a single (in)divisible good or multiple goods that are all measured under a common unit. It may therefore be formalised as a directed graph where nodes represent agents and arcs `buyer-seller' pairs. Each such pair could be characterised by a maximum exchange amount, i.e., an arc capacity. 
In addition, a social planner could assign a weight per pair to signify the, possibly differentiating, importance of the corresponding exchange (these are the ``welfare judgments" of \cite{Biro20}). For example, the social planner could wish to set unitary weights in order to signal that the social optimum is just the maximum total exchange, or set radically varying weights in order to prioritise exchanges involving socially vulnerable agents.

Each agent is indifferent regarding the goods she offers (i.e., her endowment) and also indifferent among goods offered by the same agent but expresses her strict preferences over the agents she finds acceptable to receive goods from. This is quite realistic in the sense that an individual entering a time sharing for the service she may offer has no preference of who is receiving it, as long as she gets a service she likes, and she reasonably distinguishes the most preferred persons offering it. Let us express this also in the classical house-allocation context of Shapley and Scarf \cite{Shaple74}: each agent is endowed by property rights in the form of time slots, during which she may live in one or more houses, and has a strict ordering over owners offering houses that she prefers more than hers (among which she is indifferent). In debt terms, this applies in an inverse manner: although one would happily receive any money owned to her, being indifferent on which debtor pays which amount, she must comply to an ordering regarding the debtee to be paid first (e.g., the tax office) if a relevant debt-relief cycle is available. 

Contrary to the typical housing market or the settings generalising it (e.g., \cite{Cechlarova14}), a solution to our problem is not a matching but a set of cycles in the aforementioned directed graph. 
As the cycles forming a balanced exchange are not necessarily arc-disjoint, we propose a modified TTC mechanism that finds a \emph{Pareto optimal balanced exchange}; if all arc capacities are one, it resembles the augmented top-trading-cycles (ATTC) procedure in \cite{Fujita18}. Pareto optimality is defined in terms of lexicographic preferences, as in \cite{Cechlarova14}; i.e., an exchange is Pareto optimal if there is no other exchange in which some agent gets slightly more from a more preferable agent but no agent gets something worse.  

Let us discuss relevant studies that have appeared recently. A quite similar problem is discussed in \cite{Klijn17}, i.e., a market in which each agent is endowed with multiple units of an indivisible and agent-specific good and whose outcome is a circulation defining a balanced exchange of goods. This apparently coincides with the indivisible variant of our setting, studied under a different focus: the work in \cite{Klijn17} claims that, under the thereby-called ``general capacity configurations", there is no circulation rule that satisfies individual rationality, Pareto efficiency, and strategy-proofness (we refer to \cite{Manjunath21} for relevant definitions). Another relevant setting is discussed in \cite{Manjunath21}, where each agent classifies goods as desirable or not; under this dichotomy restriction, mechanisms accomplishing all three properties are plausible. A model close to the one of \cite{Manjunath21} appears in \cite{Andersson21}, where the maximum size solution is shown to be also Pareto optimal because each agent ranks her acceptable bundles only according to their size (i.e., the total size of acceptable services or money or housing contracts she receives). Our last link is to \cite{Abbassi15}, where an agent has a large utility loss if giving away more goods than she gets and any acceptable (i.e., \emph{individually rational}) exchange (not necessarily balanced) can be viewed as a collection of directed cycles in the agent-goods bipartite graph. Motivated by kidney exchange, the results in \cite{Abbassi15} reveal how a bound on the number of these cycles imposes a limit on the approximability of a max-size exchange by any randomised mechanism that also satisfies strategy-proofness.  

Let us now discuss our contribution and how our results exploit ideas from the literature. As already mentioned, we draw inspiration from the extensive literature on kidney exchange (see \cite{Biro21} and references therein) and from the literature generalising the Shapley-Scarf market in ``many-to-many" markets of indivisible goods \cite{Cechlarova14}. The former motivates our simple variation of TTC that is specifically tailored to balanced exchanges, discussed in Section \ref{Section:TTC}. The latter motivates our characterisation of Pareto optimal exchanges as  \emph{maximal}, \emph{trade-in free} and 
\emph{coalition-free}, presented in Section \ref{Section:Char}, thus adding to the problems for which Pareto optimality can be characterised formally. Based on that characterisation, we show in Section \ref{Section:Recognise} how to identify whether a give exchange is Pareto optimal in polynomial time by transforming it to an instance of \cite{Cechlarova14}. Most importantly, the same approach allows us to obtain a Pareto optimal balanced exchange by iteratively improving any balanced exchange.

Perhaps the least obvious literature finding we employ comes from early work on integral dicycle packings \cite{Nutov95}. A dicycle packing of a directed graph is a set of arc-disjoint cycles, i.e., in our terms an exchange when all arcs have capacity $1$ or, equivalently, when each `buyer-seller' pair may exchange up to one unit. Therefore, we could validly name our setting as \emph{dicycle packing with preferences and non-unitary arc capacities.}
Although the dicycle packing problem problem is known to be $\mathcal{NP}$-hard in its integral form, its fractional version is not. By generalising some results of \cite{Nutov95}, we show how to find a max-weight fractional dicycle packing and thus a max-weight exchange in polynomial time for divisible goods under arbitrary arc capacities. This approach finds a max-weight Pareto optimal exchange if agents are indifferent not only regarding their endowment but also the goods they receive (as in \cite{Manjunath21}). 

In the presence of preferences, we show why this approach also outputs a max-weight Pareto optimal exchange if arc weights are non-increasing with respect to an agent's preference list; although we call the weights ``preference-concordant" in the terminology of \cite{Faenza21}, these are also called ``aligned interests" in \cite{Biro20}. Concordance could be considered applicable in the case where every `buyer-seller' pair agrees also on a cardinal utility that must, however, comply with the ordinal preferences of the buyer. The social planner could then wish to maximise social welfare under this valuation of bilateral transfers of goods. Alternatively, the social planner could keep the privilege of defining this valuation but in a way that does not antagonise the preferences submitted by the agents. That could become applicable, for example, in the setting of workers exchanging shifts, as described in \cite{Andersson21}. What we show here is that under concordancy there is a welfare-maximizing allocation that is Pareto efficient, in slight contrast to \cite{Biro20} where (under aligned interests) any welfare-maximizing allocation is Pareto efficient.

All results on max-weight Pareto optimal balanced exchange are presented in Section \ref{Section:Max}. The technical derivation of all the above requires some notation, discussed in Section \ref{Section:Notation} and certain definitions and interim results presented in subsequent sections. For ease of presentation, we show our results for divisible goods and simple graphs and then explain in Section \ref{Section:Conclude} that all results up to Section \ref{Section:Recognise} hold also for indivisible goods and graphs with parallel edges. Some open questions are also discussed in that concluding section.

\section{Definitions and Notation} \label{Section:Notation}

A Pareto optimal balanced exchange instance $(PX)$ can be defined as a capacitated directed graph $D=(V,A,c)$ where $c:A \rightarrow \mathbb{R}^{+}$ and every agent $v\in V$ has a strict preference list that orders the agents $u$ such that $(v,u)\in A$.
Given a set of cycles (i.e., a \emph{packing}) $\mathcal{C}$ and an arc $e\in A$, let $\mathcal{C}_{e}$ be the subset of $\mathcal{C}$ such that $e$ is an arc of $C$ for every cycle  $C\in \mathcal{C}_{e}$. 
 Let also a function $f: \mathcal{C}\rightarrow\mathbb{R}^{+}$ be an  \emph{exchange function} if  $\sum_{C\in \mathcal{C}_{e}}f(C)\leq c(e)$ for every $e\in A$.
We denote by $f(e)$ the quantity $\sum_{C\in \mathcal{C}_{e}}f(C)$.
Then, an \emph{exchange} in a $PX$ instance is a pair $(\mathcal{C},f)$ where $\mathcal{C}$ is a packing in $D=(V,A,c)$ and $f$ is an exchange function.

Observe that an agent $v$ in a $PX$ instance having either $d^{-}(v)=0$ or $d^{+}(v)=0$ can be removed from $V$ because she cannot be part of a cycle. These removals can be done recursively, hence we may assume without loss of generality that $d^{-}(v)\neq 0$,  $d^{+}(v)\neq 0$ for all $v \in V.$

Let $c^{-}(v),c^{+}(v)$ be the sum of the capacity of incoming and outgoing edges of $v$ respectively. Note that $min\{c^{-}(v),c^{+}(v)\}$ is an implicit upper bound on the quantity exchanged by agent $v.$ The existence of an explicit upper bound $c_v$ for each $v\in V$ (i.e., a node capacity) can be modelled by splitting $v$ into nodes $v^+$ and $v^-$, connecting all arcs incoming to $v$ (resp. outgoing from $v$) to $v^+$ (resp. $v^-$) and placing a capacity $c_v$ on the arc $(v^+,v^-)$. Also, arc $(v^+,v^-)$ is the single arc outgoing from $v^+$ and the single arc incoming to $v^-.$ Given this transformation, we can assume hereafter that agents may or not have an upper bound on the quantity they exchange. Also, we can present our results without explicitly considering agent capacity.

As it is customary, $v_1>_v v_2$ denotes that $v_1$ precedes $v_2$ in the preference list of $v$, thus being more preferable. Let $P(v): v_1, v_2, v_3, \ldots v_z$ be the preference list of an agent $v$ and $(\mathcal{C},f)$, $(\mathcal{C'},f')$ be two exchanges.
We say that $v$ prefers $(\mathcal{C},f)$ more than $(\mathcal{C'},f')$ and denote that by $(\mathcal{C},f)>_v (\mathcal{C'},f')$, if there exists some $k\in [z]$ such that $\sum_{C\in \mathcal{C}_{(v,v_i)}}f(C)=\sum_{C\in \mathcal{C'}_{(v,v_i)}}f'(C)$ for $i=1,2,\ldots k-1$ and 
$\sum_{C\in\mathcal{C}_{(v,v_k)}}f(C)>\sum_{C\in \mathcal{C'}_{(v,v_k)}}f'(C)$.
We say that $(\mathcal{C},f)$ dominates to $(\mathcal{C'},f')$ and denote that by $(\mathcal{C},f)\geq_D(\mathcal{C'},f')$ if there is no agent $v$ such that $(\mathcal{C},f)<_v (\mathcal{C'},f')$ and there is some agent $u$ such that $(\mathcal{C},f)>_u (\mathcal{C'},f')$
An exchange $(\mathcal{C},f)$ is Pareto optimal exchange in the usual sense, i.e., if there is no exchange where an agent can be better off without another agent worsening, i.e., there is no  $(\mathcal{C'},f')$ such that $(\mathcal{C'},f')>_D(\mathcal{C},f)$.

\section{Top Trading Cycles} \label{Section:TTC}
We present a procedure analogous to the one in \cite{Fujita18} that sends the maximum possible ``flow" over all cycles showing up in each ``top trading" iteration and then reduces residual capacities and removes vertices without any remaining incoming capacity. Let us remind the reader that the graph contains only nodes corresponding to agents.

\begin{algorithm}[h!t]
	\caption{Top Trading Cycles for Balanced Exchange}\label{TTCM}
	\begin{algorithmic}[1] 
	\REQUIRE{ A directed graph with capacities $D=(V,A,c)$. Initialize $\mathcal{C}=\emptyset$, $k=1$.}
	\ENSURE{An exchange $(\mathcal{C},f)$}
	\STATE {Create a directed graph $G^{k}(V^{k},A^{k})$ where $V^{k}=V$ and $A^{k}$ contains every arc $(v,u)$ where $v,u\in V$ and $u$ is the most preferred agent in the preference list of $v$.}
	\STATE {Let $\mathcal{C}^k$ be the set of cycles of $G^{k}$. Then $\mathcal{C}\leftarrow\mathcal{C}\cup\mathcal{C}^k$. }
	\STATE {For every $C\in \mathcal{C}^k$, set $f(C)=min_{e\in C}c(e)$.}
	\STATE {For every $e\in C$, for some $C\in \mathcal{C}^k$,  $c(e)\leftarrow c(e)-f(C)$. If now $c(e)=0$, remove $e$ from $D$. }
	\STATE {Remove from $D$ repeatedly every vertex $v$ of $V$ where $d^{+}(v)=0$}
	\STATE {If $D=\emptyset$, return  $(\mathcal{C},f)$. Otherwise $k\leftarrow k+1$ and go to the next round.} 
	\end{algorithmic}
\end{algorithm}

\begin{theorem}\label{THEDE}
Algorithm \ref{TTCM} returns a Pareto optimal balanced exchange $(\mathcal{C},f)$ for a $PX$ instance.
\end{theorem}

Let us first show the following.

\begin{lemma}
In an instance $D=(V,A,c)$ of $PX$ where $c^{-}(v)=c^{+}(v)$ for every $v\in V$ there exists an exchange $(\mathcal{C},f)$ where  $f(e)=c(e)$ for every $e\in A$.
\end{lemma}

\begin{proof}
Let $C$ be an arbitrary cycle of $D$ and $e$ be an edge of minimum capacity. 
We add $C$ to $\mathcal{C}$, we reduce the capacity of edges of $C$ by $c(e)$ and we remove every edge that now has zero capacity.
Observe that $D$ retains the property $c^{-}(v)=c^{+}(v)$ for every $v\in V$.
We repeat this process until $D$ becomes acyclic. 
Now, if there is an edge that has not been removed, then there is an edge $e=(s,t)$ that has capacity $c_e>0$ and there is no outgoing edge of $t$. 
It holds for every vertex $v$ that $\sum_{(u,v)\in A}f((u,v))=\sum_{(v,z)\in A}f((v,z))$.
Thus, $c^{-}(t)=c_e +\sum_{(u,t)\in A}f((u,t))=c_e+\sum_{(v,t)\in A}f((v,t))>\sum_{(v,t)\in A}f((v,t))=c^{+}(t)$.
This is a contradiction because $c^{-}(t)=c^{+}(t)$.
Hence, there is no edge of non-zero capacity when $D$ becomes acyclic and this concludes the proof.
\end{proof}

We can now show the main proof of this section.

\begin{proof}[Proof of Theorem \ref{THEDE}]
It is easy to see that the output  $(\mathcal{C},f)$ of Algorithm \ref{TTCM} is an exchange satisfying $\sum_{C\in \mathcal{C}_{e}}f(C)\leq c(e)$ for every $e\in A$. Moreover, $\mathcal{C}$ is maximal, as the existence of a cycle $C^{*}$ after reducing the capacity of the arcs in cycles of $\mathcal{C}$ would mean that $D$ still contained all vertices of $C^{*}.$

Assume that $(\mathcal{C},f)$ is not Pareto optimal. 
Then there is an exchange $(\mathcal{C'},f')$ such that  $(\mathcal{C'},f')>_D (\mathcal{C},f)$. If $f'(e)\geq f(e)$ for every $e\in A$ because of Pareto domination, there exist an arc $e$ where $f'(e)> f(e)$.
Let  $D_{\mathcal{C'}}=(V_{\mathcal{C'}},A_{\mathcal{C'}},f'(e)-f(e))$  where every arc of zero capacity is removed. As both exchanges $(\mathcal{C'},f')$ and $(\mathcal{C},f)$ are balanced, it holds that $c^{-}(v)=c^{+}(v)$ for every $v\in V_{\mathcal{C'}},$ hence the graph $D_{C'}$ cannot be acyclic.
This yields that there is an arc such that $f(e)>f'(e)$.

Let $f_r(e)$ be the value of $f(e)$ after round $r$ and $k$ be the earliest round where for an arc $(v,u)$ it holds that $f_k((v,u))>f'((v,u))$.
If $(\mathcal{C},f)>_v (\mathcal{C'},f'),$ there is a contradiction to $(\mathcal{C'},f')>_D (\mathcal{C},f')$. Thus, $(\mathcal{C'},f')>_v (\mathcal{C},f)$ and there exists an arc $(v,t)$ where $f'((v,t))>f((v,t))$ and $t>_v u$. Let $C'$ be a cycle of $\mathcal{C'}$ where $(v,t)$ belongs to.
As $t>_v u$ and $f'((v,t))>f((v,t)),$ $t$ must have already been removed from $D$ at round $k$ of Algorithm \ref{TTCM}; otherwise $u$ would not be the most preferred agent for $v$ at this round.
This implies that there is an arc $e$ of $C'$ where $f_{k-1}(e)=c(e)$.

Let $(a,b)$ be the first arc with this property in the directed path $C'\setminus\{(v,t)\},$ i.e., the vertices of the path from $t$ to $a$ in $C'$ are among the ones removed in round $k$, including $a$ itself; also let $(z,a)$ be an arc of $C'$. Now consider the graph $D^{*}=(V,A,f'-f_{k-1})$ and remove from it all arcs with zero capacity. It holds that $c^{-}(u)=c^{+}(u)$ for every vertex $u$ of $D^{*}$ and that $(z,a)$ is an arc of this graph so there is an arc $(a,s)$ in $D^{*}$. As $c((a,s))-f_{k-1}((a,s))>0,$ this contradicts the fact that $a$ has been removed at round $k$. 
Hence, $(\mathcal{C},f)$ is Pareto optimal.
\end{proof}

\section{Characterization} \label{Section:Char}

In this section we present a characterization of Pareto optimal balanced exchanges, analogous to the one introduced for Pareto optimal matchings \cite{Cechlarova14}.
Let $(\mathcal{C},f)$ be an exchange in a directed graph $D=(V,A,c).$ We denote as $A_{\mathcal{C}}$ the arcs of the graph over which goods are exchanged in $\mathcal{C}$, i.e., $ A_{\mathcal{C}}=\{e\in A: f(e)>0\}.$ 
Let also $D^{\prime}$ denote the graph with residual capacities, i.e., $D^{\prime}=(V,A^{\prime})$ where $A^{\prime}=\{e\in A: c(e)-f(e)>0\}$. Now we can present the definitions of maximal, trade-in-free and coalition-free in $PX$.

\begin{definition} \label{Def:Conditions} An exchange $(\mathcal{C},f)$ is
\begin{itemize}
    \item[(i)]  \emph{maximal} if $D^{\prime}$ is acyclic,
    \item[(ii)] \emph{trade-in-free} if for every $(v,u)\in  A_{\mathcal{C}}$ there is no $(v,t,\ldots,u)$ path in $D^{\prime}$ and $t>_v u$, and 
    \item[(iii)] \emph{coalition-free} if for every $(v_1,u_1),(v_2,u_2),\ldots,(v_k,u_k) \in  A_{\mathcal{C}}$ there are no paths $(v_1,t_1,\ldots,u_2)$,$(v_2,t_2,\ldots,u_3)$, $\ldots, (v_k,t_k,\ldots,u_1)$ in $D^{\prime}$ such that $t_i>_{v_i} u_i$ for every $i\in[k].$
\end{itemize}
\end{definition}

We can now present the main proof, whose technical details are quite elaborate, thus being followed by an illustrative example.

\begin{theorem} \label{Thm:Charact}
An exchange $(\mathcal{C},f)$ is Pareto optimal if and only if it is maximal, trade-in-free and coalition-free.
\end{theorem}

\begin{proof}
It is rather direct to check that a Pareto optimal exchange it is maximal, trade-in-free and coalition-free, as otherwise it could be improved and thus dominated. 

Conversely, assume that an exchange $(\mathcal{C},f)$ is maximal, trade-in-free and coalition-free but not Pareto optimal. 
Then, there is an exchange $(\mathcal{C'},f')$ such that $(\mathcal{C'},f')>_D (\mathcal{C},f)$.
Consider now that if there exists an exchange that dominates  $(\mathcal{C},f)$, then there must be a maximal exchange that dominates $(\mathcal{C},f)$; thus, without loss of generality, $(\mathcal{C'},f')$ is maximal.

Let $f^{\ast}(e)=min\{f(e),f'(e)\}$ for every $e\in A.$ 
Let also $D_1=(V_1,A_1)$, $D_2=(V_2,A_2)$ where $A_1=\{e\in A_{\mathcal{C}}: f(e)-f^{\ast}(e)>0\}$ $(A_2=\{e\in A_{\mathcal{C'}}: f'(e)-f^{\ast}(e)>0\})$ and $V_1 (V_2)$ are the vertices that are incident to the arcs of $A_1 (A_2)$. We denote by $d_1^+(v)$ and $d_1^-(v)$  $(d_2^+(v)$ and $d_2^-(v))$ the out-degree and the in-degree of a vertex $v$ respectively  in $D_1$ $(D_2)$.

Because of maximality of $(\mathcal{C},f)$ and $(\mathcal{C'},f')$, $D_1$ and $D_2$ are both acyclic: e.g., if there were a cycle $C$ in $D_2$, then  $c(e)-f(e)=c(e)-f^{\ast}(e)\geq f'(e)-f^{\ast}(e)>0$ for every $e\in C$, hence $C$ would also be a cycle in $D^{\prime}$ and $(\mathcal{C},f)$ would not be maximal. Also note that $d_1^+(v)>0$ implies $d_2^+(v)>0$ for any $v$, as otherwise $(\mathcal{C},f)>_v (\mathcal{C'},f')$, a contradiction to the assumption that $(\mathcal{C'},f')>_D (\mathcal{C},f)$. 

To complete the proof given these observations, we need to construct another directed graph $D_0(V_0,A_0)$ with the property that, if $(\mathcal{C},f)$ complies to Definition \ref{Def:Conditions} but is dominated, all vertices in $V_0$ have an outgoing arc and that in turn implies that there exists a cycle in $D_0$ that leads to a Pareto improvement of $(\mathcal{C},f)$. 

\noindent The vertex set $V_0$ is obtained as follows:

\begin{itemize}
    \item[(i)] if $d_1^-(v)=0$ and $d_1^+(v)>0$ , $v$ is added to $V_0$;
    \item[(ii)] if $d_1^+(v)=0$ and $d_1^-(v)>0$, $v$ is added to $V_0$;
    \item[(iii)] if $d_1^-(v)>0$ and $d_1^+(v)>0$, two vertices $v_{in}$ and $v_{out}$ are added to $v_0$; and,
    \item[(iv)] if $v\notin V_1$ and $d_2^+(v)>0$, $v$ is added to $V_0$.
\end{itemize}
The arc set $A_0$ is constructed as follows.
\begin{itemize}
    \item[(i)] If $(v,u)\in A_1$, then $(u,v)$ or $(u_{in},v)$ or $(u,v_{out})$ or $(u_{in},v_{out})$ is added to $A_0$, depending on whether $v$ or $v_{out}$ and $u$ or $u_{in}$ appears in $V_0$; we denote these arcs by $A_r$ since they are reversed.
    \item[(ii)] If $(v,u)\in A_2$, $v\in V_1$, $d_1^+(v)>0$ and $u$ is the most preferred agent for $v$ in $D_2$, then $(v,u)$ or $(v_{out},u)$ or $(v,u_{in})$ or $(v_{out},u_{in})$ is added to $A_0,$ depending on whether $v$ or $v_{out}$ and $u$ or $u_{in}$ appears in $V_0$.
    \item[(iii)] If $(v,u)\in A_2$ and $v\notin V_1$, $(v,u)$ or $(v,u_{in})$ is added to $A_0,$ depending on whether $u_{in}$ or $u$ appears in $V_0$.
\end{itemize}
Let us explain why, by construction, every vertex in $D_0$ has out-degree at least 1.
\begin{itemize}
    \item[(i)] If $d_1^-(v)=0$ and $d_1^+(v)>0$, then $d_2^+(v)>0$ and an arc of type (ii) starting from $v$ belongs to $A_0$.
    \item[(ii)] If $d_1^+(v)=0$ and $d_1^-(v)>0$, there exists an arc $(u,v)\in A_1$ so that $(v,u)\in A_0$ (type (i)).
     \item[(iii)] If $d_1^-(v)>0$ and $d_1^+(v)>0$, then $d_2^+(v)>0$ so that there exist an arc $(v_{out},u)$ in $A_0$ (type (ii)); moreover, if  $(z,v)\in A_1$ then $(v_{in},z)\in A_0$ (type (i)).
     \item[(iv)] If $v\notin V_1$ and $d_2^+(v)>0$, then an arc of type (iii) starting from $v$  belongs to $A_0$.
\end{itemize}
The fact that every vertex in $D_0$ has out-degree at least 1 implies that there is a directed cycle $C$ in $D_0$. If $C$ has no arcs of $A_r$, then there exists also a cycle in $D_2$. Thus, $(\mathcal{C},f)$ is not maximal and this completes the proof. 

Otherwise, let $C=(u_1,v_1,\ldots,u_2,v_2,\ldots,u_k,v_k,\ldots, u_1)$ where $(u_i,v_i)\in A_r$ for $i\in[k].$ By construction of $D_0$, there are no consecutive arcs of $A_r$ in $C$. 
Moreover, if $(v_i,t_i)$ is an arc of $C$, then either $t_i=u_{i+1}$ or $t_i\in V_2\setminus V_1$ and $t_i$ is the most preferred agent for $v_i$ in $D_2$.
Now, in what follows, if a vertex $z_i$ of $C$ is of type $z_{in}$ or $z_{out}$, let $z_i$ be the origin vertex of $D_1$. 
It holds that $t_i>_{v_i} u_i$ for every $i\in[k]$ because $t_i$ is the most preferred agent for $v_i$  and $(\mathcal{C'},f')>_D(\mathcal{C},f)$. 
This implies that if $k=1$, there exists an arc $(v_1,u_1)\in A_{\mathcal{C}}$ and a  path $(v_1, t_1,\ldots, u_1)$ in $D^{\prime}$ where $t_1>_{v_1} u_1$, i.e., $(\mathcal{C},f)$ is not trade-in-free.
If $k>1$, there exist $(v_1,u_1),(v_2,u_2),\ldots,(v_k,u_k) \in A_{\mathcal{C}}$ and paths $(v_1,t_1,\ldots,u_2),(v_2,t_2,\ldots,u_3),\ldots, (v_k,t_k,\ldots,u_1)$ in $D^{\prime}$ and $t_i>_{v_i} u_i$ for every $i\in[k],$ But then, $(\mathcal{C},f)$ is not coalition-free and this completes the proof.
\end{proof}

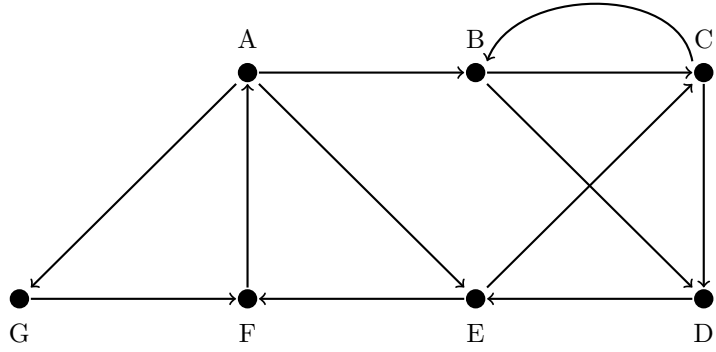
\begin{figure}[tbp]
\begin{center}
\begin{tikzpicture}[thick,scale=1.5]
\draw[fill=black] (0,0) circle (0.5ex);
\draw[fill=black] (2,0) circle (0.5ex);
\draw[fill=black] (0,2) circle (0.5ex);
\draw[fill=black] (4,0) circle (0.5ex);
\draw[fill=black] (2,2) circle (0.5ex);
\draw[fill=black] (4,2) circle (0.5ex);
\draw[fill=black] (-2,0) circle (0.5ex);

\node at (0,-0.3){F};
\node at (2,-0.3){E};
\node at (4,-0.3){D};
\node at (0,2.3){A};
\node at (2,2.3){B};
\node at (4,2.3){C};
\node at (-2,-0.3){G};

\draw[->] (0,0.1)--(0,1.9);
\draw[->] (0.1,2)--(1.9,2);
\draw[->] (2.1,2)--(3.9,2);
\draw[->] (4,1.9)--(4,0.1);
\draw[->] (3.9,0)--(2.1,0);
\draw[->] (1.9,0)--(0.1,0);
\draw[->] (0.1,1.9)--(1.9,0.1);
\draw[->] (2.1,1.9)--(3.9,0.1);
\draw[->] (2.1,0.1)--(3.9,1.9);
\draw[->] (-1.9,0)--(-0.1,0);
\draw[->] (-0.1,1.9)--(-1.9,0.1);
\path[->,out=100,in=70] (3.9,2.1) edge (2.1,2.1);
\end{tikzpicture}
\end{center}
\caption{The directed graph of Example \ref{ex1}}
\label{Fig1}
\end{figure}

\begin{example}\label{ex1}
In the graph $D=(V,A,c)$ in Figure \ref{Fig1}, let $c((F,A))=2$ and $c(e)=1$ for every other arc of $A.$ The preference lists of the agents are as follows:
\begin{align*}
    &P(A): G, E, B& &P(E): C, F \\
    &P(B): D, C& &P(F): A \\
    &P(C): B, D& &P(G):F\\
    &P(D): E & 
\end{align*}

Consider the exchange $(\mathcal{C},f)$ that contains the cycles $C_0=(A,B,C,D,E,F,A)$ and $C_1=(A,G,F,A),$ where $f(C_0)=f(C_1)=1$. Consider also the exchange $(\mathcal{C'},f')$ that contains the cycles $C_1$, $C_2=(A,E,F,A)$ and $C_3=(B,D,E,C,B)$ where $f'(C_1)=f'(C_2)=f'(C_3)=1$. Observe that $(\mathcal{C'},f')>_D (\mathcal{C},f)$. The graph $D_1$ of the proof of Theorem \ref{Thm:Charact} is the path $(A,B,C,D);$ and the graph $D_2$ is the path $(A,E,C,B,D)$ i.e., both are acyclic. 

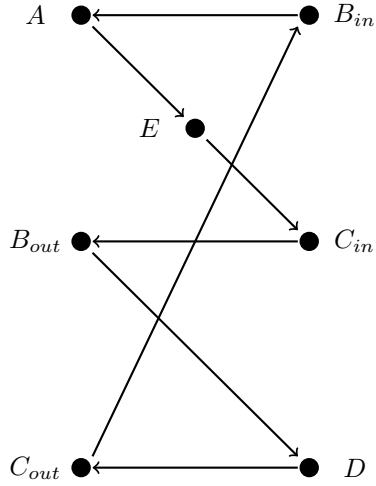
\begin{figure}[tbp]
\begin{center}
\begin{tikzpicture}[thick,scale=1.5]
\draw[fill=black] (0,0) circle (0.5ex);
\draw[fill=black] (2,0) circle (0.5ex);
\draw[fill=black] (0,2) circle (0.5ex);
\draw[fill=black] (0,4) circle (0.5ex);
\draw[fill=black] (2,2) circle (0.5ex);
\draw[fill=black] (2,4) circle (0.5ex);
\draw[fill=black] (1,3) circle (0.5ex);

\node at (-0.4,0){$C_{out}$};
\node at (-0.4,4){$A$};
\node at (2.4,4){$B_{in}$};
\node at (-0.4,2){$B_{out}$};
\node at (2.4,2){$C_{in}$};
\node at (2.4,0){$D$};
\node at (0.6,3){$E$};

\draw[->] (1.9,4)--(0.1,4);
\draw[->] (1.9,2)--(0.1,2);
\draw[->] (1.9,0)--(0.1,0);
\draw[->] (0.1,3.9)--(0.9,3.1);
\draw[->] (1.1,2.9)--(1.9,2.1);
\draw[->] (0.1,1.9)--(1.9,0.1);
\draw[->] (0.1,0.1)--(1.9,3.9);

\end{tikzpicture}
\end{center}
\caption{The graph $D_0$ of Example \ref{ex1}.}
\label{Fig2}
\end{figure}

The graph $D_0$ of the example appears in Figure \ref{Fig2}. The existence of the cycle $(A,E,C_{in},B_{out},D,C_{out},B_{in},A)$ implies that $(\mathcal{C},f)$ is not Pareto optimal. The coalition replaces the arcs $(A,B),(B,C)$ and $(C,D)$ of $(\mathcal{C},f),$ by the paths $(A,E,C),(B,D)$ and $(C,B)$ which is an improvement for $(\mathcal{C},f)$. $\blacksquare$
\end{example}

\section{Recognition} \label{Section:Recognise}

Theorem \ref{Thm:Charact} allows us to obtain a polynomial-time method that recognizes whether a given balanced exchange in an instance of $PX$ is Pareto optimal. This method is based on constructing a one-to-many Pareto optimal matching instance in which one can check if an appropriate matching is Pareto optimal \cite{Cechlarova14}.

Given a directed graph with capacities $D=(V,A,c)$ and an exchange $(\mathcal{C},f)$, let us remind the reader that $A_{\mathcal{C}}$ is the set of arcs that belong to a cycle of $\mathcal{C}$ and $D^{\prime}=(V,A^{\prime})$ has $A^{\prime}=\{e\in A: c(e)-f(e)>0\}$. Let also $q_v$ be the number of arcs that belong to a cycle of $\mathcal{C}$ and whose ending vertex is $v.$ 

We construct a one-to-many matching instance with one-sided preferences $\mathcal{I_{\mathcal{C}}}$, where we denote the respective bipartite graph by $G_{\mathcal{C}}=(S_{\mathcal{C}}\cup T_{\mathcal{C}},E_{\mathcal{C}}$), as follows:

\begin{itemize}
    \item[(i)] The set of agents is $S_{\mathcal{C}}=\{v_u: (v,u)\in A_{\mathcal{C}})\}$.
    \item[(ii)] The set of objects is $T_{\mathcal{C}}=\{v: v\in V \text{ and } q_v>0\}$.
    \item[(iii)] The quota of an object $v$ is $q_v$ and the quota of each agent is $1.$
    \item[(iv)] For every agent $v_u$ let $D^{\prime}_{vu}$ be the graph obtained by $D^{\prime}$ after removing the arcs $(v,t)$ where $u>_v t$. The preference list of the agent $v_u$ contains, in an arbitrary order, the objects whose respective vertices are reachable from $v$ in $D^{\prime}_{vu}$ and $u$ in the last position of the list. \end{itemize}
\addtocounter{example}{-1}
\begin{example} (cont.)
Consider the exchange $(\mathcal{C},f)$ where $\mathcal{C}$ contains the cycles $C_0=(A,B,C,D,E,F,A)$, $C_1=(A,G,F,A)$ and has $f(C_0)=f(C_1)=1$. The instance $\mathcal{I_{\mathcal{C}}}$ contains the objects $A,B,C,D,E,G$ with quota $1$ and $F$ with quota $2$. 
The agents and their preference lists may be as follows:

\begin{align*}
    &P(A_B): E,C,B,D,B& &P(D_E): E\\ 
    &P(A_G): G & &P(E_F): C,B,D,F\\
    &P(B_C): D, C & &P(F_A):A\\
    &P(C_D): B,D,D &  &P(G_F):F  
\end{align*}
\end{example} 

There are two critical aspects regarding our construction, related to maximality and trade-in-freeness. We explain these in detail, although they impose no barrier to our method, as both properties can be checked in polynomial time before constructing the instance of one-to-many matching with one-sided preferences.

First, if an object $v$ appears in the preference list of $v_u$, this implies that ($\mathcal{C},f)$ is not maximal. However, if there is no appearance of an object $v$ in the list of an agent $v_u$ this does not exclude the non-maximality of ($\mathcal{C},f);$ such a ``weird" situation occurs if there exists a cycle in $D^{\prime}$ such that every vertex of the cycle does not belong to any cycle of $\mathcal{C}$. Remind that by Definition \ref{Def:Conditions}i we can easily answer if  $(\mathcal{C},f)$ is maximal by checking if $D'$ is acyclic.

Secondly, an object $u$ may appear twice in the preference list of $v_u,$ i.e., $G_{\mathcal{C}}$ may have parallel edges. In the above example, $B$ and $D$ appear twice in the preference lists of $A_B$ and $C_D,$ respectively. But then, one can observe that $(\mathcal{C},f)$ is not trade-in-free because the arc $(A,B)$ can be replaced by the (improving) path $(A,E,C,B)$ and $(C,D)$ by $(C,B,D)$. 
The following lemma connects Definition \ref{Def:Conditions}ii with the existence of parallel edges in $G_{\mathcal{C}}$.

\begin{lemma}\label{apptwice}
$(\mathcal{C},f)$ is trade-in-free if and only if $G_{\mathcal{C}}$ has no parallel edges.  
\end{lemma}

\begin{proof}
By construction of $G_{\mathcal{C}},$ only an edge of the form $(v_u,u)$ where $(v,u)\in A_{\mathcal{C}}$ could be a parallel edge. Hence, assume that $\mathcal{C}$ is trade-in-free and let an arc $(v,u)$ be such an arc of $A_{\mathcal{C}}$. Then, $v_u\in S_{\mathcal{C}}$ and $u$ is the last entry in the preference list of $v_u$. 
As $(\mathcal{C},f)$ is trade-in-free and $(v,u)\in A_{\mathcal{C}},$ there is no $(v,t,\ldots,u)$ path in $D^{\prime}$ and $t>_v u$ i.e., $u$ is not reachable by $v$ in $D^{\prime}_{vu}$ so $(v,u)$ is a simple edge of $G_{\mathcal{C}}$.

Conversely, assume that there are no parallel edges in $G_{\mathcal{C}}$.
Let $(v_u,u)$ be an edge of $G_{\mathcal{C}}$ where $(v,u)\in A_{\mathcal{C}}$ may be a parallel edge. As $(v_u,u)$ is a simple edge, then $u$ is not reachable by $v$ in $D^{\prime}_{vu}$. Thus, there is no $(v,t,\ldots,u)$ path in $D^{\prime}$ where $t>_v u$ and this implies that $(\mathcal{C},f)$ is trade-in-free.
\end{proof}

We can recognize polynomially if $(\mathcal{C},f)$ is maximal and trade-in-free by checking if $D^{\prime}$ is maximal and $G_{\mathcal{C}}$ has parallel edges, respectively. Thus we may safely assume that we construct $I _{\mathcal{C}}$ starting with a maximal and trade-in-free exchange. That is, in what follows, the instance $I _{\mathcal{C}}$ is used to check only if $(\mathcal{C},f)$ is coalition-free. 

The definitions of a maximal, trade-in-free and coalition-free bipartite matching are as in \cite{Cechlarova14}. Define $M_{\mathcal{C}}=\{(v_u,u): v_u\in S_{\mathcal{C}} \},$ which is unique as there are no parallel edges. In $M_{\mathcal{C}}$, every agent is matched once and every object $v$ is matched exactly $q_v$ times.  This implies that $M_{\mathcal{C}}$ is trade-in-free and maximal in  $\mathcal{I_{\mathcal{C}}}$. 

\addtocounter{example}{-1}
\begin{example} (cont.)
Considering exchange $(\mathcal{C},f),$ a coalition with respect to $M_{\mathcal{C}}$ is $\{(A_B,B),(B_C,C),(C_D,D)\}$. Observe that this coalition regarding $M_{\mathcal{C}}$ corresponds to the coalition regarding $(\mathcal{C}, f)$ that replaces $(A,B),(B,C)$ and $(C,D)$ with $(A,E,C),(B,D)$ and $(C,B)$, respectively. Therefore, using the instance $I _{\mathcal{C}}$ we can detect that $(\mathcal{C},f)$ is not Pareto optimal and, in addition, improve it to obtain the exchange $(\mathcal{C^{\prime}},f^{\prime}).$  $\blacksquare$
\end{example}
Let us recall from  that a coalition in a bipartite graph with one-sided preferences $\{(a_0,c_0),(a_1,c_1),\ldots,(a_k,c_k)\}$ is related to the cycle $(a_0,c_1,a_1,c_2,a_2,\ldots,c_k,a_k,c_0,a_0)$. 
The following theorem completes the recognition of Pareto optimality for a balanced exchange.

\begin{theorem} \label{thm:recognize}
Let $(\mathcal{C},f)$ be a maximal and trade-in-free exchange in an instance of $PX$. Then $(\mathcal{C},f)$ is Pareto optimal if and only if the matching $M_{\mathcal{C}}$ is Pareto optimal in the instance $I_{\mathcal{C}}$.
\end{theorem}

\begin{proof}
Firstly, assume a matching $M_{\mathcal{C}}$ defined in $I_{\mathcal{C}}$, which is not a Pareto optimal matching in $I_{\mathcal{C}}$. As $M_{\mathcal{C}}$ is maximal and trade-in-free, this implies that $M_{\mathcal{C}}$ is not coalition-free (by \cite[Theorem 1]{Cechlarova14}).
Let $R=\{(v_{u_1},u_1),(v_{u_2},u_2),\ldots,(v_{u_k},u_k)\}$ be a coalition with respect to $M_{\mathcal{C}}$. Thus, $u_{i+1}>_{v_{u_i}} u_i$ for every $i=1,2,\ldots,k$ i.e, there exist a path $(v_i,t_i,\ldots,u_{i+1})$ in $D^{\prime}$ such that $t_i>_{v_i} u_i$.
The coalition is related to the cycle $C=(v_{u_1},u_2,v_{u_2},u_3,\ldots,v_{u_{k-1}},u_k,v_{u_k},u_1,v_{u_1})$. 
Recall that the edges of $G_{\mathcal{C}}$ represent an arc or a direct path in $D$. Given these, we construct a directed closed walk by replacing the edges in $C$ with the paths of $\mathcal{C},$ where the arcs of $D$ that are represented by edges of the coalition $R$ are reversed. Hence, this closed walk has the form: $$W=(v_1,t_1,\ldots,u_2,v_2,t_2,\ldots,u_3,\ldots, v_{k-1},t_{k-1},\ldots,u_k,v_{k},t_{k},\ldots,u_1,v_1)$$ 
where $k\geq 2$ because $R$ contains at least two pairs. After removing the reversed arcs of $\mathcal{C}$, a set of directed paths are obtained i.e., $(v_1,t_1,\ldots,u_2),$ $(v_2,t_2,\ldots u_3), \ldots,(v_k,t_k,\ldots ,u_1)$.
As $t_{i}>_{v_{i}} u_{i}$ for $j=1,2,\ldots,k$, these paths comprise a coalition in which the arcs $(u_1,v_1),(u_2,v_2), \ldots,(u_k,v_k)$ of $\mathcal{C}$ have been replaced by the aforementioned paths. This contradicts the assumption that $(\mathcal{C},f)$ is not coalition-free and  concludes that, if $M_{\mathcal{C}}$ is not a Pareto optimal matching in $I_{\mathcal{C}}$ then $(\mathcal{C},f)$ is not a Pareto optimal exchange.
    

Now, assume that $(\mathcal{C},f)$ is not Pareto optimal. By the hypothesis of the current Theorem and Theorem \ref{Thm:Charact}, it can only be that $(\mathcal{C},f)$ is not coalition-free. Thus, there are $(v_1,u_1),(v_2,u_2),\ldots,(v_k,u_k) \in A_{\mathcal{C}}$ such that there exist paths $(v_1,t_1,\ldots,u_2)$, $(v_2,t_2,\ldots,u_3),$ $\ldots, (v_k,t_k,\ldots,u_1)$ in $D^{\prime}$ and $t_i>_{v_i} u_i$ for every $i\in[k].$ 
In $I_{\mathcal{C}}$ we denote $v_{u_i}$ the agent obtained by the arc $(v_i,u_i)$ of $A_{\mathcal{C}}$. The object $u_{i+1}$ (modulo $k$) belongs in the preference list of $v_{u_i}$ because of the path $(v_i,t_i,\ldots,u_{i+1})$ and as $t_i>_{v_i} u_i$ then $v_{u_i}$ prefers $u_{i+1}$ to $u_i$. Thus, $(v_{u_1},u_1),(v_{u_2},u_2),\ldots,(v_{u_k},u_k)$ is a coalition with respect to $M_{\mathcal{C}}$ in $I_{\mathcal{C}}$. This proves that $M_{\mathcal{C}}$ is not a Pareto optimal matching in $I_{\mathcal{C}}$ and completes the proof.
\end{proof}

By \cite[Corollary 1]{Cechlarova14}, the Pareto optimality of a bipartite matching with one-sided preferences can be determined in polynomial time. This, in conjunction with Theorem \ref{thm:recognize}, implies the same in our setting. If a given exchange is not Pareto optimal, the recognition method yields an improvement that leads to an exchange that dominates the initial one.
Repeating this, a Pareto optimal exchange is obtained after a number of iterations that is also polynomial because, at each iteration, at least two agents receive goods from a more preferred agent. Overall, we have the following.

\begin{corollary} \label{Cor:PX}
Given any maximal and trade-in-free exchange, a Pareto optimal exchange can be obtained in polynomial time.
\end{corollary}

Observe that Corollary \ref{Cor:PX} allows for multiple methods - i.e., mechanisms - for obtaining a Pareto optimal exchange, possibly quite different from TTC. This idea is exploited in the next section.

\section{Max-weight Pareto Optimal Exchange} \label{Section:Max}

As discussed in the introductory section, each arc can be accompanied by a weight that signifies the social importance per unit of the associated exchange. A more general weighted version would be to assume a weight function $w:\mathcal{P}\rightarrow \mathbb{R^{+}}$ where $\mathcal{P}$ is the set of all directed cycles in $D$. Then, the problem of finding a max-weight Pareto optimal balanced exchange is denoted by $PX(w)$. Although the complexity of $PX(w)$ is unknown, we present two cases where $PX(w)$ can be solved polynomially, both having potential applicability.

\subsection{Indifference} \label{Sec:Indiff}

If the agent expresses no preference over the goods she receives or the agents she gets something from, we have a version of our problem that is quite close to that of \cite{Andersson21,Manjunath21}. If, in addition, all cycles have weight $1,$ $PX(w)$ is equivalent to the  fractional dicycle packing problem which can be solved polynomially, as its linear programming formulation suffices to contain only inequalities associated with cycles of length at most $3$ \cite[Theorem 14]{Nutov95}. 

We prove a similar result in the case where the weight of a cycle is additive, i.e., it is the sum of the weights per arc. That is, let $l:A\rightarrow \mathbb{R^{+}} \cup \{0 \}$ where $l(e)=0$ if and only if $e$ is of the form $(v^+, v^-),$ i.e., it has been introduced to accommodate the capacity of agent $v,$ as discussed in Section 
\ref{Section:Notation}. Let $w(C)=\sum_{e\in C}l(e)$ for every cycle $C$ in $D$ (if all arc weights are $1$ then $w$ is the length of the cycle). Assume that $D=(V,A,c)$ is a complete directed graph by adding arcs of zero capacity and weight for every $v,u\in V$ and $(v,u)\notin A$. $PX(w)$ can be described as a linear program as follows.
\begin{align*}
 &\text{ max }  \sum_{C\in D}y(C)w(C) \\
 & \sum_{C\in \mathcal{C}_{e}}y(C)\leq c(e) \text{ for every arc } e\in A \\
 & y\geq 0
\end{align*}

The dual of the above linear program is the following:
\begin{align} \nonumber
 &\text{ min }  \sum_{e\in A}c(e)x(e) \\ \label{dual}
 & \sum_{e\in C}x(e)\geq w(C) \text{ for every cycle } C\text{ in } D \\ 
 & x\geq 0 \nonumber
\end{align}

To modify \cite[Theorem 14]{Nutov95} for our purposes, we need the following lemma.

\begin{lemma}\label{setofcycles}
For any $v,u\in V$ and $(u,v),(v,u)\in A$, let $C_{uv}$ and $C_{vu}$ be two cycles such that $(u,v)\in C_{uv}$ and $(v,u)\in C_{vu}$ and let also  $C=(C_{uv}\setminus(u,v))\cup (C_{vu}\setminus(v,u))$. Then, there is a set of cycles $C_1,\ldots , C_k$ such that $C=\bigcup_{i=1}^{k} C_i$ and any arc $(x,y)\in C$ appears in exactly two such cycles if $(x,y)\in (C_{uv}\setminus(u,v))\cap (C_{vu}\setminus(v,u))$ and a single such cycle otherwise.
\end{lemma}

\begin{proof}
Let $C_{uv}=(u,v,x_1,\ldots,x_r,u)$ and $C_{vu}=(v,u,y_1,\ldots,y_p,v)$. Then  $(v,x_1,\ldots,x_r,u,y_1,\ldots,y_p,v)$ is a closed walk and it is obvious that an arc of $C=(C_{uv}\setminus(u,v))\cup (C_{vu}\setminus(v,u))$ appears in the walk once if it belongs to exactly one of $C_{uv}\setminus(u,v)$ and  $C_{vu}\setminus(v,u)$, and twice if it belongs to $(C_{uv}\setminus(u,v))\cap (C_{vu}\setminus(v,u))$. 
If the walk is a cycle, the proof is complete. Otherwise, there exist $j\in[r],m\in[p]$ such that $x_j=y_m$. Let $j$ be the minimum number with this property.
Then, $C_1=(v,x_1,\ldots,x_{j-1},x_j=y_m,y_{m+1},\ldots,y_p,v)$ is a cycle.
Now $C=C_1\cup C'$ where $C'=(x_j,x_{j+1},\ldots,x_r,u,y_1,\ldots , y_m=x_j)$ is a closed walk. 
Then, $C'$ is either a cycle or let $j'$ be the minimum number such that $j'>j$ and $x_{j'}=y_{m'}$ where $y_{m'}\in C'$. As previously, a cycle $C_2$ is obtained. Repeating this procedure decomposes the closed walk into a set of cycles $C_1,\ldots , C_k$; as no arc is deleted, this completes the proof.
 \end{proof}

We can now present our main result that directly shows the solvability of $PX(w)$ in polynomial time under indifference and additive weights.

\begin{theorem} \label{Thm:Indiff}
If $w(C)=\sum_{e\in C}l(e)$, the linear program
\begin{align}
\nonumber
 &\text{ min }  \sum_{e\in A}c(e)x(e) \\ \label{dualpol}
 & \sum_{e\in C}x(e)\geq \sum _{e\in C}l(e) \text{ for every cycle  } C\text{ of length 3 in } D \\ \nonumber
  & \sum_{e\in C}x(e)= \sum _{e\in C}l(e) \text{ for every cycle  } C\text{ of length 2 in D }\\ \nonumber
 & x\geq 0
\end{align}
is equivalent to (\ref{dual}), i.e., (\ref{dual}) can be solved by minimizing a linear function over a polynomial number of constraints.
\end{theorem}

\begin{proof}
Let $x^{*}$ be an optimal solution of (\ref{dual}) such that if $y\leq x^{*}$ and  $y\neq x^{*}$, then $y$ is not a solution of (\ref{dual}).
This implies that for every $(u,v)\in A$ $$x_{uv}^{*}+x_{vu}^{*}=l((u,v))+l((v,u)).$$ 
The edges $(u,v)$ and $(v,u)$ form a cycle of length 2, hence $x_{uv}^{*}+x_{vu}^{*}\geq l((u,v))+l((v,u))$. 

Assume that for some $u,v\in V$ $x_{uv}^{*}+x_{vu}^{*}> l((u,v))+l((v,u))$.
Because of the minimality of $x^{*}$, one can derive two cycles $C_{uv}$ and $C_{vu}$ such that $(u,v)\in C_{uv}$,   $(v,u)\in C_{vu}$, $x^{*}(C_{uv})=w_{C_{uv}}$ and $x^{*}(C_{uv})=w_{C_{vu}}$. Otherwise, if for example for every cycle $C_{uv}$ that contains the arc $(u,v)$ it holds that $x^{*}(C_{uv})>w_{C_{uv}}$, this implies that $x_{uv}^{*}$ can be reduced, a contradiction to $x^{*}$ being a minimal optimal solution.

Now, let $C=(C_{uv}\setminus(u,v))\cup (C_{vu}\setminus(v,u))$. Thus,

$$ x^{*}(C)=x^{*}(C_{uv})+x^{*}(C_{vu})-(x_{uv}^{*}+x_{vu}^{*})=w_{C_{uv}}+w_{C_{vu}}-(x_{uv}^{*}+x_{vu}^{*})<w_{C_{uv}}+w_{C_{vu}}-l((u,v))+l((v,u)).$$
 By Lemma \ref{setofcycles}, $C=\bigcup_{i=1}^{k} C_i$ and every arc has the same number of appearances in $C$ and in $\bigcup_{i=1}^{k} C_i$. Moreover it holds that $\sum_{i=1}^{k}w_{C_i}=w_{C_{uv}}+w_{C_{vu}}-l((u,v))+l((v,u))$. This leads to a contradiction because
$$x^{*}(C)=\sum_{i=1}^{k}x^{*}(C_i)\geq \sum_{i=1}^{k}w_{C_i}=w_{C_{uv}}+w_{C_{vu}}-l((u,v))+l((v,u)).$$
Hence, for  every $(u,v)\in A$, $x_{uv}^{*}+x_{vu}^{*}=l((u,v))+l((v,u)).$ Then (\ref{dual}) is equivalent to 
\begin{align*}
 &\text{ min }  \sum_{e\in A}w(e)x(e) \\
 & \sum_{e\in C}x(e)\geq \sum_{e\in C} l(e)\text{ for every cycle  } C\text{ in } D \\
  & \sum_{e\in C}x(e)=  \sum_{e\in C} l(e) \text{ for every cycle  } C\text{ of length 2 in D }\\
 & x\geq 0
\end{align*}

It can be shown by induction on the length of the directed cycles, that if $x$ is a feasible
solution of the system defined in (\ref{dualpol}), 
then $x(C)\geq w(C)$ for any cycle $C$ in $D$. 
Assume that this holds for every cycle of length at most  $m$. We prove that this holds for a cycle $C_{m+1}$ of length $m+1$. 
Let $C_{m+1}=(v_1,v_2,\ldots,v_{m+1},v_1)$ and  $v_i$ and $v_j$ be two non-consecutive vertices of $C_{m+1}$ and without loss of generality $i<j$.
Then, $C_1=(v_1,v_2,\ldots,v_i,v_j,v_{j+1},\ldots v_{m+1},v_1)$ and $C_2=(v_i,v_{i+1},\ldots,v_j,v_i)$ are cycles of length at most $m$ and the sum of their length is $l_{C_1}+l_{C_2}=l_{C_{m+1}}+l((u,v))+l((v,u))$. 
Hence, by the induction hypothesis $x(C_1)\geq w_{C_1}$ and $x(C_2)\geq w_{C_2}.$
Now,
$$x(C_{m+1})=x(C_1)+x(C_2)-x(v_i,v_j)-x(v_j,v_i)\geq w_{C_1}+w_{C_2}-l((u,v))-l((v,u))=\sum_{e\in C_{m+1}} l(e)$$
 This yields that the system defined above and the one defined in (\ref{dualpol}) are the same, and the
proof is complete.
\end{proof}

\subsection{Concordance} 

As in \cite{Faenza21}, we say that $l:A\rightarrow \mathbb{R}^{+}$ is \emph{preference-concordant}  if, for each $(x,y),(x,y') \in A$, $y>_x y'$ whenever $l((x,y)) \geq l((x,y'))$.
Moreover if $l((x,y)) > l((x,y'))$ for every $(x,y),(x,y') \in A$ where $y>_x y'$ we say that $l$ is \emph{strictly preference-concordant}. 

We call a $PX(w)$ instance $\mathcal{I}$ preference-concordant if $w(C)=\sum_{e\in C} l(e)$ and $l$ is preference-concordant. To find a max-weight Pareto optimal exchange, we first compute a maximum exchange $(\mathcal{C},f)$ of $\mathcal{I}$ as in Section \ref{Sec:Indiff}. Such an exchange is maximal because arc weights are positive, thus cycle weights are also positive.

Let us remind the reader that the objective function of the linear programming is $\sum_{C\in D}y(C)w(C)$. As $(\mathcal{C},f)$ is the exchange that maximizes this function, it holds that $\sum_{C\in \mathcal{C}}f(C)w(C)=\sum_{e\in A_{\mathcal{C}}} f(e)l(e)$.

\begin{lemma} \label{tradmax}
In an preference-concordant $PX(w)$ instance, a max-weight exchange is trade-in-free.
\end{lemma}

\begin{proof}
Assume that $(\mathcal{C},f)$ is a max-weight exchange that is not trade-in-free. 
Then there exist an arc $(u,v)\in A_{\mathcal{C}}$ and a path $P=(v,t_0,t_1,\ldots t_k,u)$  in $D^{\prime}$ where $t>_v u$. Let $f_t=min\{f((v,u)),c((v,t_0))-f((v,t_0)),c((t_0,t_1))-f((t_0,t_1)),\ldots,c((t_k,u))-f((t_k,u))\}$; $f_t$ is the maximum amount that can be transferred from $(v,u)$ to $P$.
Let also $(\mathcal{C'},f')$ be the exchange obtained by transferring the amount $f_t$ from $(v,u)$ to the path $(v,t_0,t_1,\ldots t_k,u)$. 
As $l$ is preference-concordant and $t>_u v$, then $l((v,t))\geq l((v,u))$ so that $l(P)>l((u,v))$ and 
$$\sum_{C\in \mathcal{C'}}f'(C)w(C)=\sum_{e\in A_{\mathcal{C'}}} f'(e)l(e)=(\sum_{e\in A_{\mathcal{C}}} f(e)l(e))+f_t(l(P)-l((v,u)))>\sum_{e\in A_{\mathcal{C}}} f(e)l(e)=\sum_{C\in \mathcal{C}}f(C)w(C).$$
This contradicts that $(\mathcal{C},f)$ is a maximum exchange, thus $(\mathcal{C},f)$ is trade-in-free.
\end{proof}

In a coalition, let $e$ be an arc that belongs to $\{(v_1,u_1),(v_2,u_2),\ldots,(v_k,u_k)\}$ such that $f(e)\leq f((v_i,u_i))$ for every $i=1,2,\ldots,k$. Let $A_P$ be the set of arcs that belongs to the paths $(v_1,t_1,\ldots,u_2),(v_2,t_2,\ldots,u_3),\ldots, (v_k,t_k,\ldots,u_1)$. These paths are not edge-disjoint. Let $m_a$ be the number of appearances of arc $a$ in $A_P$. Let $e'$ be an arc of $\mathcal{P}$ such that $\frac{c(e')-f(e')}{m_{e'}}\leq \frac{c(a)-f(a)}{m_a}$ for every $a\in \mathcal{P}.$ Let $f_c=min\{f(e),\frac{c(e')-f(e')}{m_{e'}}\}$. That is, given a coalition, $f_c$ is the maximum  amount that can be transferred from the arcs  $(v_1,u_1),(v_2,u_2),\ldots,(v_k,u_k)$ to the paths  $(v_1,t_1,\ldots,u_2),(v_2,t_2,\ldots,u_3),\ldots, (v_k,t_k,\ldots,u_1)$. 

\begin{lemma}\label{coalmax}
In a preference-concordant $PX(w)$ instance, let $(\mathcal{C},f)$ and $(\mathcal{C'},f')$ be two exchanges such that $(\mathcal{C'},f')$ is obtained by  $(\mathcal{C},f)$ after transferring the maximum possible amount along a coalition. Then, $\sum_{C\in \mathcal{C'}}f'(C)w(C) \geq \sum_{C\in \mathcal{C}}f(C)w(C)$.
\end{lemma}

\begin{proof}
Let $(\mathcal{C'},f')$ be obtained after replacing in $(\mathcal{C},f)$ the arcs  $(v_1,u_1),(v_2,u_2),\ldots,(v_k,u_k)$  of $ A_{\mathcal{C}}$ by the paths $P_1=(v_1,t_1,\ldots,u_2),P_2=(v_2,t_2,\ldots,u_3),\ldots, P_k=(v_k,t_k,\ldots,u_1)$ of $D^{\prime}$ where $t_i>_{v_i} u_i$ for every $i=1,2,\ldots,k$.
Let $f_c$ be  the maximum possible amount of this coalition, defined as in the above discussion.
Since $l$ is preference-concordant and $t_i>_{u_i} v_i$, then $l((v_i,t_i))\geq l((v_i,u_i))$ so that $l(P_i)\geq l((v_i,u_i))$ for every $i=1,2,\ldots,k$.
It holds that $$\sum_{C\in \mathcal{C'}}f'(C)w(C)=\sum_{e\in A_{\mathcal{C'}}} f'(e)l(e)=(\sum_{e\in A_{\mathcal{C}}} f(e)l(e))+f_c\sum_{i=1}^{k}(l(P_i)-l((v_i,u_i)))  \geq \sum_{e\in A_{\mathcal{C}}} f(e)l(e)=\sum_{C\in \mathcal{C}}f(C)l_C.$$
\end{proof}

We can now show how to compute a max-weight exchange in a a preference-concordant $PX(w)$ instance. 

\begin{theorem}
In a preference-concordant $PX(w)$ instance, a max-weight Pareto optimal exchange can be computed in polynomial time.
\end{theorem}
\begin{proof}
We start with a max-weight exchange produced through Theorem \ref{Thm:Indiff}. Such an exchange $(\mathcal{C},f)$ is maximal and trade-in-free by Lemma \ref{tradmax}, thus it is Pareto optimal unless there is a coalition. In such a coalition, every path is of length $1$ and $l((u_i,v_i))=l((u_i,v_{i+1}))$ for every $i=1,2,\ldots,k$. Using Corollary \ref{Cor:PX}, we can improve it to a Pareto optimal exchange in polynomial time, while Lemma \ref{coalmax} yields that these improvements do not compromise the weight. 
\end{proof}

A simple example shows that max-weight balanced exchanges are not necessarily Pareto optimal under concordance, in contrast to the case of ``aligned interests" in \cite{Biro20}.

\begin{example}\label{exmaxnotPar}
Consider the graph of Figure \ref{maxnotPar} and let $A>_B C$, $C>_D A$, $c\equiv 1$, $l((B,A))=l((B,C))$ and  $l((D,A))=l((D,C))$. The balanced exchange $(\mathcal{C},f)$ that contains $C_0=(A,B,C,D,A)$ where $f(C_0)=1$ is maximum but not Pareto optimal, as it is dominated by the exchange  $(\mathcal{C}',f')$ that contains the cycles $C_1=(A,B,A)$ and $C_2=(C,D,C)$ where $f'(C_1)=f'(C_2)=1.$ That is, there exists a coalition in  $(\mathcal{C},f)$ that replaces the arcs $(B,C)$ and $(D,A)$ by the arcs $(B,A)$ and $(D,C)$.
\begin{figure}[tbp]
\begin{center}
\begin{tikzpicture}[thick,scale=1.5]
\draw[fill=black] (0,0) circle (0.5ex);
\draw[fill=black] (2,0) circle (0.5ex);
\draw[fill=black] (0,2) circle (0.5ex);
\draw[fill=black] (2,2) circle (0.5ex);

\node at (-0.3,0){D};
\node at (2.3,0){C};
\node at (-0.3,2){A};
\node at (2.3,2){B};

\draw[->] (0,0.1)--(0,1.9);
\draw[->] (0.1,2)--(1.9,2);
\draw[->] (1.9,0)--(0.1,0);
\draw[->] (2,1.9)--(2,0.1);
\path[->,out=100,in=70] (2,2.1) edge (0,2.1);
\path[->,out=280,in=250] (0,-0.1) edge (2,-0.1);
\end{tikzpicture}
\end{center}
\caption{The directed graph of Example \ref{exmaxnotPar}}
\label{maxnotPar}
\end{figure}
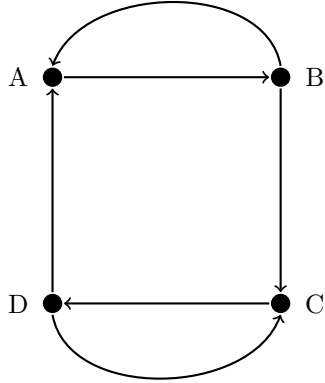
\end{example}

In the proof of Lemma \ref{coalmax}, it is easy to see that if $l$ is strictly preference-concordant then $\sum_{C\in \mathcal{C'}}f'(C)w(C) > \sum_{C\in \mathcal{C}}f(C)w(C)$. Therefore, any maximum exchange is maximal, trade-in-free and coalition-free and, hence, Pareto optimal.

Lifting concordance no longer guarantees that there is a maximum exchange that is Pareto optimal. 
\begin{example}\label{ex-conc}
        In the graph of Figure \ref{concord}, let  $B>_A D$, $c\equiv 1$ and $l((A,B))<l((A,D))$. Then the single Pareto optimal exchange is the cycle $C_1=(A,B,C,A),$ while the max-weight exchange is the cycle $C_2=(A,D,C,A);$ $f=1$ in both cases. 
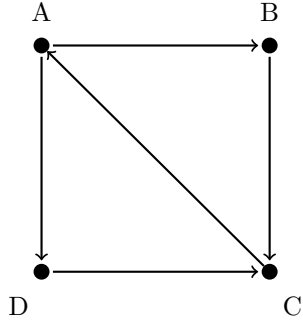
\begin{figure}[tbp]
        \centering

\begin{tikzpicture}[thick,scale=1.5]
\draw[fill=black] (0,0) circle (0.4ex);
\draw[fill=black] (2,0) circle (0.4ex);
\draw[fill=black] (0,2) circle (0.4ex);
\draw[fill=black] (2,2) circle (0.4ex);

\node at (-0.2,-0.3){D};
\node at (2.2,-0.3){C};
\node at (0,2.3){A};
\node at (2,2.3){B};

\draw[->] (0,1.9)--(0,0.1);
\draw[->] (0.1,2)--(1.9,2);
\draw[->] (0.1,0)--(1.9,0);
\draw[->] (2,1.9)--(2,0.1);
\draw[->] (1.95,0.05)--(0.05,1.95);

\end{tikzpicture}

\caption{The graph of Example \ref{ex-conc} }
\label{concord}
\end{figure}
\end{example}

\section{Implications and further work} \label{Section:Conclude}
We studied balanced exchanges assuming a single divisible good or different agent-specific goods, thus having each agent's preferences expressed over other agents rather than goods. Our results include a simple TTC variant, a characterisation result leading to a polytime recognition scheme and a special case in which a max-weight Pareto optimal balanced exchange can be computed in polynomial time.

It would be interesting to see if there are conditions other than concordancy under which the (fractional) max-weight exchange is Pareto optimal. The complexity of finding a max-weight Pareto optimal exchange would be worth investigating too.

Our results move beyond our setting as some of them can also be transferred  to the case of indivisible good(s), possibly offered in multiple copies. This implies that the capacity function $c$ is an integer function and in an exchange $(\mathcal{C},f)$ the function $f$ must be also integer. It is obvious that the TTC mechanism can be used because if $c$ is an integer function, then the obtained by TTC mechanism function $f$ is also integer. The characterisation and recognition of Pareto optimal balanced exchanges apply also under indivisibility. Notice that, in this case, the paths in Definition \ref{Def:Conditions}iii must be arc-disjoint as the transferred amount of the coalition cannot be divided.

\begin{lemma}
If an integral exchange contains a coalition, then it also contains a coalition with arc-disjoint paths.
\end{lemma}
\begin{proof}
A directed closed walk contains a directed trail. Consider a closed walk $W'=(z_0,z_1,\ldots,z_r,z_0)$ contains some arcs multiple times and let $(z_i,z_{i+1})=(z_j,z_{j+1})$ for some $i,j$ where $i<j$. Assume that there are no $k,l$ such that $k<i$ and $l>j$ where $(z_i,z_{i+1})=(z_k,z_{k+1})$ and $(z_j,z_{j+1})=(z_l,z_{l+1})$. Thus, in the closed walk $(z_0,z_1,\ldots,z_{i-1},z_{i}=z_j,z_{i+1}=z_{j+1},z_{j+2},\ldots,z_r,z_0)$ it holds that the arc $(z_i,z_{i+1})$ appears once. We repeat this until a trail $T'$ is obtained. 
Let $(z_i,z_{i+1},z_{i+2})$ be a part of $W'$ and $(z_i,z_{i+1})$ be an arc that appears once in $W'$. Observe that if $(z_i,z_{i+1})$ is an arc of $T'$ then the next arc in $T'$ is $(z_{i+1},z_{i+2})$. 
Let in a coalition the arcs $(v_1,u_1),(v_2,u_2),\ldots,(v_k,u_k)$ be replaced by  $(v_1,t_1,\ldots,u_2),(v_2,t_2,\ldots,u_3), \ldots, (v_k,t_k,\ldots,u_1)$.
The arcs $(v_1,u_1),(v_2,u_2),\ldots,(v_k,u_k)$ are reversed. Then the coalition is related with the closed walk
$W=(v_1,t_1,\ldots,u_2,v_2,t_2,\ldots,u_3,\ldots,v_k,t_k,\ldots,u_1,v_1)$. The obtained trail by the walk $W$ leads to a coalition with arc-disjoint paths. 
\end{proof}

Unfortunately, our results on max-weight Pareto optimal exchange cannot be extended because finding an integral dicycle packing is $\mathcal{NP}$-hard. This however does not directly imply that the integral $PX(w)$ for additive weights and/or under preference concordance is $\mathcal{NP}$-hard.


Moving from indivisibility to multiplicity of goods, let us observe that our TTC variant works correctly even if parallel edges - representing different goods - are present. It would be reasonable to examine whether the same holds also for our characterisation and recognition results.

\end{document}